1. Title page:

**Title:** The potential of artificial intelligence for achieving healthy and sustainable societies

Name: B. Sirmacek
Email: bsirmacek@gmail.com
University/affiliation: Smart Cities, School of Creative Technologies, Saxion University of Applied Sciences, Enschede, The Netherlands

Name: S. Gupta
Email: shivam.gupta@uni-bonn.de
University/affiliation: Bonn Alliance for Sustainability Research, University of Bonn, Germany

Name: F. Mallor
Email: mallor@kth.se
University/affiliation: FLOW, Engineering Mechanics, KTH Royal Institute of Technology, Stockholm, Sweden

Name: H. Azizpour
Email: azizpour@kth.se
University/affiliation: Division of Robotics, Perception, and Learning, KTH Royal Institute of Technology, Stockholm, Sweden

Name: Y. Ban
Email: yifang@kth.se
University/affiliation: Division of Geoinformatics, KTH Royal Institute of Technology, Stockholm, Sweden

Name: H. Eivazi
Email: hamidre@kth.se
University/affiliation: FLOW, Engineering Mechanics, KTH Royal Institute of Technology, Stockholm, Sweden

Name: H. Fang

Email: hfang@kth.se
University/affiliation: Division of Robotics, Perception, and Learning, KTH Royal Institute of Technology, Stockholm, Sweden

Name: F. Golzar
Email: fargo@kth.se
University/affiliation: Division of Energy Systems, Department of Energy Technology, KTH Royal Institute of Technology, Stockholm, Sweden
Climate Action Centre, KTH Royal Institute of Technology, Stockholm, Sweden

Name: I. Leite
Email: iolanda@kth.se
University/affiliation: Division of Robotics, Perception, and Learning, KTH Royal Institute of Technology, Stockholm, Sweden

Name: G.I. Melsion
Email: gimp@kth.se
University/affiliation: Division of Robotics, Perception, and Learning, KTH Royal Institute of Technology, Stockholm, Sweden

Name: K. Smith
Email: ksmith@kth.se
University/affiliation: Division of Robotics, Perception, and Learning, KTH Royal Institute of Technology, Stockholm, Sweden

Name: F. Fuso Nerini
Email: francesco.fusonerini@energy.kth.se
University/affiliation: Division of Energy Systems, Department of Energy Technology, KTH Royal Institute
Climate Action Centre, KTH Royal Institute of Technology, Stockholm, Sweden

Name: R. Vinuesa
Email: rvinuesa@mech.kth.se

University/affiliation: FLOW, Engineering Mechanics, KTH Royal Institute of Technology, Stockholm, Sweden

Climate Action Centre, KTH Royal Institute of Technology, Stockholm, Sweden

**Abstract**

In this chapter we extend earlier work (Vinuesa et al., Nature Communications 11, 2020)on the potential of artificial intelligence (AI) to achieve the 17 Sustainable Development Goals (SDGs) proposed by the United Nations (UN) for the 2030 Agenda. The present contribution focuses on three SDGs related to healthy and sustainable societies, i.e.SDG 3 (on good health),SDG 11 (on sustainable cities) and SDG 13 (on climate action). This chapter extends the previous study within those three goals, and goes beyond the 2030 targets. These SDGs are selected because they are closely related to the coronavirus disease 19 (COVID-19) pandemic, and also to crises like climate change, which constitute important challenges to our society.

## 1. Introduction

In the past years, driven by the increased capacity in acquisition, storage and processing of data, artificial intelligence (AI) has emerged a disruptive technology, affecting a broad scope of fields. As these capacities are only increasing, AI has cemented its impact on society as a whole, and it is therefore expected to play a crucial role in the achievement of the Sustainable Development Goals (SDGs) proposed by the United Nations (UN) [2]. As pointed out by Vinuesa et al. [163], AI can enable the achievement of 134 out of the 169 targets accompanying the SDGs. Nonetheless, AI can also act as an inhibitor of 59 of the SDG targets. Therefore, special care should be taken when deploying AI solutions at a large scale, and its impact (positive or negative) on society, economy and the environment should be carefully assessed. The on-going coronavirus-disease pandemic (COVID-19) has shown the dangers that a major crisis can have on the urban-population health. Moreover, it has been a prime example of the use of AI and big data, e.g. through the use of contact-tracing apps which have evidenced both the positives (effectiveness) and negatives (privacy, ethical issues) derived from the use of AI [145, 164]. In this regard, the current climate emergency presents itself as the next major crisis to be faced by our species. In this chapter, we focus our analysis on the impact of AI on the SDGs related to healthy and sustainable societies, i.e. SDG 3 (on good health), SDG 11 (on sustainable cities) and SDG 13 (on climate action).The chapter is structured as follows: firstly, impacts of AI adoption on health are assessed in §2. Secondly, in §3 we look at the role of AI in the achievement of sustainable cities. Then, we focus our attention to the possibilities enabled by AI when it comes to climate-action targets in §4.1. Lastly, general conclusions regarding the effect of AI on achieving healthy and sustainable societies are drawn, and an outlook is presented in §5.

## 2. Improved health through AI

### 2.1. Shortage of health-care workforce

One main challenge within the health sector is the shortage of care staff, especially in developing countries. In 2006, the World Health Organization (WHO) estimated a global shortage of 4.3 million health workforce, identifying it as a crisis [123]. Later in 2016, a WHO report projected a shortage of 18 million health workers by 2030 [125]. While a large part of the shortage concerns lack of nursing staff, the lack of enough physicians and tertiary-care staff is also remarkable, even within developed countries [107, 133], which is exacerbated when considering the capacity required for training specialists. Such a shortage is highly imbalanced against low-and middle-income (LAMIC) countries globally and rural areas within the individual countries. Furthermore, a recent Lancet report estimated 5.7 million deaths per year in LAMIC countries due to poor health care, or lack thereof [93]. In some countries, mitigating this shortage of staff would require hundreds of years given the current medical-training infrastructure. The recent report by the UN High Commission on health employment and economic growth [126] puts forward recommendations to mitigate these issues, one of which is digital transformation of healthcare services. The AAMS report [107] points to the promise of artificial intelligence (AI) to address the demand for specialists in various domains. Recent advances of AI techniques, especially deep learning [99], can help alleviate the severity of such shortage from numerous aspects, including: i) prioritizing care to patients under the limitation of resources such as care staff, medical equipment or hospital beds; ii) estimating the probability of having or risk of developing a medical condition given a patient's family history or own historical data and examinations; iii) monitoring patients and suggesting possible follow-ups, treatments, or patient's outcome based on the patient's condition, its severity, its risk of degradation, and available alternative actions; iv) more efficient and less costly education and training of additional care staff; and v) discovering more effective biomarkers and treatments.

The primary focus of AI research in medicine has, so far, been on the automatic diagnosis of diseases and conditions using electronic health records (EHR) [115, 135] and imaging data

[14,45] or risk thereof based on patient's own and family history [154] and radiological [34] or other types of imaging [16, 140]. Moreover, AI models can help identify the most appropriate course of action for possible follow-up examinations [44] and potential treatments [179]; accordingly they can predict the patient's outcome [83]. AI methods assisting the health staff with diagnosis, screening, and prognosis can lead to a relative reduction of their workload [134], furthermore, a better risk model can help prioritize patients and focus the limited resources to reduce mortality and morbidity rates. In fact, patient triaging is an active area of research for the AI research in medicine with promising results [95, 104]. While AI-assisted diagnosis, screening, prognosis, and triaging have potential for global application in the near future, there are other aspects where AI research has shown potential for a slightly more distant future. One central area is in AI assisting the training of health care staff [148] that can significantly reduce the cost of education, increase the efficiency of the trainings and crucially enhance the agility of care training programs adapting to the emerging needs [126] recommends reducing barriers to education which can be facilitated by AI. Finally, the most notable future applications of AI to help with the shortage of care staff are robotics surgery [92], discovery of more efficient and accurate biomarkers [171] and treatments [28], especially in light of AlphaFold's recent breakthrough in computational biology [85].

## 2.2. One Health and AI

Pandemics such as COVID-19, ebola, and cholera have grave consequences for health, economy, and society. Unless we understand comprehensively what causes them, they will emerge again and again. Usually, infectious diseases are often unleashed by microorganisms such as viruses and bacteria having very diverse origins. The change in land use type and surrounding ecosystems brings humans in close proximity with wild species that could transmit unknown pathogens. Thus, the possible way to prevent epidemics and pandemics is to realize the interconnection between human, animal, and environmental health, as it is covered under the One health domain. Target 3.3 of Agenda 2030 addresses aims to address concerns related to epidemics and other communicable disease. However, the real challenge is understanding the dynamics of the disease spread is how to better comprehend the vast amount of interdisciplinary data sources from the areas of interface between the health of humans, animals, plants, and the environment, fundamental to the One Health approach [31, 90]. AI is supportive of addressing multiple challenges faced by the field of One health. For instance, antimicrobial resistance (AMR) relation to infectious diseases was considered as one of the three One-Health priorities during the tripartite (FAO-OIE-WHO) meeting of 2011 [124]. Recently, algorithms helped identify an antibiotic called Halicin from a vast digital collection of pharmaceutical compounds [152]. AI is also helping in the management of multidrug-resistant by predicting infection risk, identifying the aetiology and misuse of antibiotics, and estimating the risk of emergence [13, 55]. Researchers are already applying the AI capabilities to support clinical decision-making processes, such as radiology, dermatology, pathology, and ophthalmology, improving further the one health infrastructure [51]. AI is also supporting prognosis-related applications using electronic health record-based clinical decision support[40], generating alerts by an AI model that provides an early warning. AI models are also supporting to predict deterioration and identifying possible pathogens and antibiotic susceptibility [5]. At a broader level AI is also helping to link diverse remote-sensing data sources for diverse One Health sub-domains [27, 157].

## 2.3. GeoAI for precision medicine

AI and data-science techniques are supportive of developing efficient, accurate, and productive knowledge for healthcare and medicine, also known as Health intelligence (HI) [144]. AI is aiding in multiple aspects of HI Health, such as in syndromic surveillance with social media[185], at-risk populations prediction [135], enabling m-health services [79], and medical imaging analysis [128]. Integrating multiple data health information with spatial context, Geospatial artificial intelligence (GeoAI) represents a focused application of AI within health intelligence to extract precise location-relevant information that helps in taking concrete action to improve health, and well-being [75]. GeoAI is helping to integrate mobile health (mHealth) information in precision medicine by consolidating information on exposures to environmental factors such as air, noise, luminescence, etc., with location to improve the spatio-temporal information for precision medicine [84]. GeoAI further supports precision medicine with geomedicine, a sub-domain that deals with individuals' location history for disease diagnosis and treatment [19]. GeoAI capabilities help clinicians access patients health considering crucial aspects related to ambient exposures to environmental risk factors of where they lived, worked, and traveled for tailored prevention and treatment strategies. However, methodological challenges concerning the limited availability of labeled training datasets, scarce standards and protocols for integrating diverse data sources, and data privacy concerns need to be recognized for sustainable development.

## 2.4. Ethical and societal considerations

Previous studies have placed SDG 3 (on good health and well-being) in a unique position for AI to make an impact on [65,127]. Vinuesa et al. [163] found SDG 3 to be the goal where AI could have the least inhibitory effect, while showing a great potential to bring several of its targets forward. However, the socio-ethical context of how and where AI technology is used in health-care systems could result in an increase of inequalities between different population groups and-nations, hence hindering its capabilities to act as enabler of other SDGs e.g. SDG 10 on reducing inequalities, and/or progressing at a lower rate among population groups with e.g. lower AI literacy or ability to access to the technology itself [47, 168]. Fenech and Buston [47] investigated the perception of healthcare professionals, technologists, ethicists and patients about the challenges of introducing AI into healthcare systems and they found that ethical, social and political questions were raised across various aspects. From the change within the relationships between patients and healthcare professionals and their acceptance of AI in a health setting, to the implications of collaboration between public and private sectors and its regulation, while also going through the concerns of responsible data handling, transparency and its impacts on existing health inequalities. In line with the shift of the patient-clinician relationship, there is concern regarding the responsibility that healthcare workers will hold on educating the patient about the complexities of AI and its possible shortcomings, or even in which cases would be required to notify that AI is being used at all [53]. This could also come with an additional burden for health professionals that may be required to get specialised training with the latest advances in the field (currently a relevant matter of global discussion [30, 174]), which might create a rebound effect increasing their workload – becoming more evident in developing countries due to the current digital divide [184]. In this sense, understanding of healthcare workforce's perceptions of AI is crucial fora successful implementation and deployment of new systems [147], and also to increasing their trust on AI since this may be an issue causing hesitancy in their predisposition to use the technology. Hence, ultimately providing the right tool to foster a partnership between the clinicians and AI [161].

A major concern regarding the applicability of AI into healthcare systems is the collection, handling, and use of patient data, due to the fact that these systems require high amounts of personal health information to accomplish accurate results. In their review of the ethics literature in AI applications for good health, Murphy et al. [118] found that three out of four common themes being discussed by researchers revolved around the impacts of collecting and using data– namely, privacy and security, trust in AI applications, and adverse consequences of bias. People may be reluctant with sharing health information if a secure and reliable process is not in place to ensure privacy and an ethical use of it [164]. For instance, the risk of the data being hacked is a reason why patients may decide not moving their health information to a digital, cloud-based format, but also the potential misinformation about the ways and application where their data is going to be used [109, 118]. The possibility of the same data collected for healthcare systems to be valuable for unrelated applications of different corporations or governments is an important threat to users trust in these systems, such as psychological data potentially being used for assessing prisoners' recidivism or by insurance companies to rate their investment risk [109]. Then, a strong legal framework is necessary to ensure transparency also on the boundaries to which personal medical records could be used. An important example is how to handle the proportionality of data sharing that is required to advance development of AI systems [129], e.g., the amount or how long are different stakeholders able to hold onto the data, as in the case of DeepMind given access to 1.6 billion UK citizens' medical records indefinitely to improve an application for managing acute kidney injuries [132].

In terms of the problematic of bias, there are different aspects that may play a role when it comes to the entire life cycle and development of AI systems. The lack of diversity in gender, ethnicity and socio-economic background of the people developing AI solutions is an important factor to address the bias in AI research and development [173], together with the fact of close assessment of the data used for training of the systems. There have been several examples of AI algorithms that produce biased results because of certain groups of the population being under-represented in the datasets [22, 121, 122, 188, 189], which can increase mistrust in these systems. Moreover, the interpretation given to the dataset in use may also cause inadvertent biased predictions when the proxies that drive an algorithmic decision are unfair towards a certain group. This has been the case for an AI system used in the US to determine patients that will require complex and intensive future health care needs [121]. A black patient with the same risk scores a white one would be less likely to be enrolled into the program because there exists a historic difference in the cost of health care between ethnicities, which is the variable used for the prediction – cost is unbiased from an underlying data point of view, yet an imperfect proxy that fails in taking into account the important social perspective causing biased predictions [121]. In this sense, there is an ongoing debate of the accountability and liability related to the recommendations and decisions made by AI systems [109, 118, 162], and how to determine who should be held responsible in the case of bad consequences of their outcomes – a topic discussed in computer ethics research for decades [35].

## 3. Towards sustainable cities with the help of AI

Rapid urbanization poses significant social and environmental challenges, including sprawling informal settlements, increased pollution, urban heat islands, loss of biodiversity and ecosystem services, as well as making cities more vulnerable to disasters. Therefore, timely and accurate information on urban areas and their changing patterns is of crucial importance to support planning sustainable and climate-resilient cities and communities. With its synoptic view, large area coverage at regular revisits, satellite remote sensing has been playing a crucial role in mapping the spatial patterns of urban areas and monitoring their temporal changing trajectories. Earth observation (EO) satellites are now acquiring massive amount of satellite imagery with higher spatial resolution and frequent temporal coverage. These EO big data represent a great opportunity to develop innovative methodologies for urban mapping and continuous change detection. The main challenge used to be the lack of robust and automated processing methods to extract valuable information from the huge volume of EO data. Advanced mathematical methods with development of AI architectures and processing platforms allowed rapid extraction of reliable information from such big data. In the following subsections we discuss the most frequently used AI architectures and applications in order to bring solutions for the sustainable development of cities aiming towards climate adaptation.

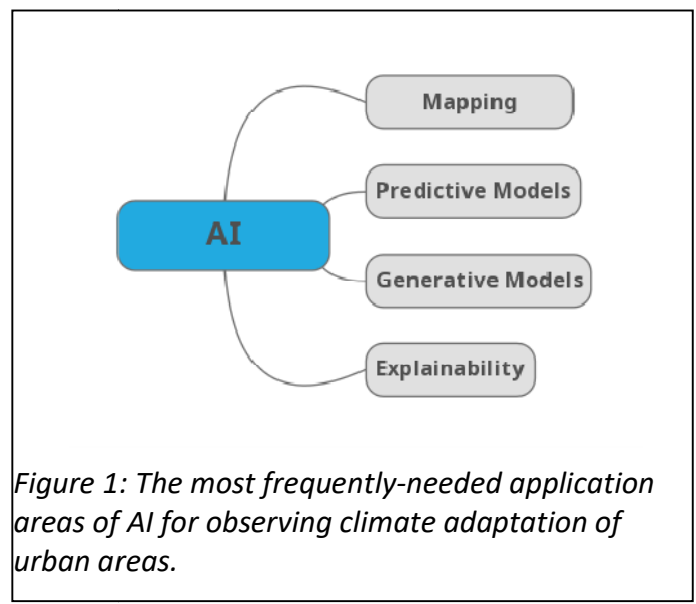

*Figure 1: The most frequently-needed application areas of AI for observing climate adaptation of urban areas.*

### 3.1. AI for extracting climate-related indicators from cities

Urban areas are responsible for creating heat islands which cause very high ecological stress to the environment. This stress is not only caused for the urban areas, but also for the surrounding areas even if they are not occupied by human activities. This phenomenon is known as urban-heat-island (UHI) effect [110]. For accurate identification of the heat-island sources and the environmental changes within the urban areas, many IT-infrastructured (also known as 'smart') cities, have been putting efforts and resources to collect a good amount of data which might be helpful to understand the stress factors. Thus in many smart cities, citizens, government institutions, industry and scientists share data for the benefit of all (this relation is also known as 'Quintuple Helix model' [23]). Obviously, this leads to a great amount of data collection and the need for machine-learning (ML) or artificial-intelligence (AI) models which can be used for understanding the climate impacts and develop preparedness aligned with the SDGs. AI can enable various applications to support cities. Figure 1 shows the AI-based applications which may have the most potential to provide immediate benefit for climate-related observation and preparedness. There might be even more applications which can be achieved by AI algorithms, however herein we keep our focus on mapping, predictive modeling, generative modeling and explainability applications. In the following subsections, we will discuss each of these application areas in detail.

### 3.1.1. Mapping

For observing climate adaptation of large areas in sustainable manner, the most frequently used data comes from satellite imaging. Satellite remote sensing allows us to collect data and information about earth surface, oceans and the atmosphere at several spatio-temporal scales in a timely, regular and accurate manner [182]. Satellite data help us understand the climate system in generally and it might help to identify ways to adapt urban regions for the drastic impacts of the climate change. Various organisations like NASA, NOAA, ESA, JAXA use satellite data to monitor greenhouse gases concentration in atmosphere, weather patterns, vegetation health, melting of glaciers and polar ice, bleaching of coral reefs, ocean acidification, changes in wildlife migratory patterns, and many other environment indicators. When it comes to urban areas, such maps are useful to identify changes of the urban structures, vegetation, agriculture, air quality, surface temperature and soon,. Besides satellite imagery, it is also possible to collect data about urban regions using airborne sensors and other in-situ internet-of-things (IoT) sensors. Higher-resolution data achieved from such resources might enrich the information given in the maps as well. In Figure 2, we provided some of the valuable information which can be extracted and visualized in urban maps to observe their climate adaption. AI algorithms can help with the following areas:

- automatic identification and mapping of trees [130]
- early recognition of forest fires [187]
- measuring the earth surface temperature and predicting the urban heat island impacts [87]
- detecting roads and traffic density [18]
- creating 3D building models [175]
- understanding agriculture health for food security [96]
- understanding soil health and properties [117]
- observing water qualities [156]
- observing air pollution [10]
- predicting and mapping air flow [60]
- analyzing and merging IoT data [3]

There are, of course, more application areas where AI algorithms help with creating maps which are useful to understand the sustainable development needs and to provide opportunity to create further action plans. However, it is challenging to address all of them, therefore we kept our focus on the most-frequently focused application areas.

### 3.1.2. Predictive models

One of the most impactful features of AI is its capability to help with building effective predictive modeling algorithms. AI models can allow fitting predictive models for data which have high numbers of degrees of freedom and exhibit non-linearities [67]. Long short-term memory (LSTM) for instance (an artificial recurrent neural network architecture used in the field of predictive modeling) is able to store information over a period of time. In other words, LSTM networks have a memory capacity for both long- and short-term periods of data. This characteristic is extremely useful when we deal with time-series data. LSTM models can decide which time series information to remember and which information to discard while creating the predictive model and making future predictions. Thus, such AI models behave robuster than the earlier mathematical models [71]. Scientists have found opportunities to use such advanced AI models for observing climate adaptation of urban areas. Advancements of AI, therefore, allowed prediction of future heat island impacts [87], water security [167] and further climate-adaptation goals for the future.

fig2

### 3.1.3. Generative models

AI might yield further applications for SDG 11 with its generalization capabilities. To this end, a special AI structure called generative adversarial network (GAN) learns deep representations without extensively annotated training data. They achieve this by deriving backpropagation signals through a competitive process involving a pair of networks. The representations that can be learned by GANs may be used in a variety of applications, including image synthesis, semantic image editing, style transfer, image super-resolution and classification [33, 80]. In the context of climate adaptation of urban areas, generative models are frequently used for data augmentation, which helps to create labels for other deep learning applications [74]. They are also found useful for creating pixel-based and accurate semantic segmentations without seeing so many examples [29] and for creating super-resolution images from course satellite-based observations [172].

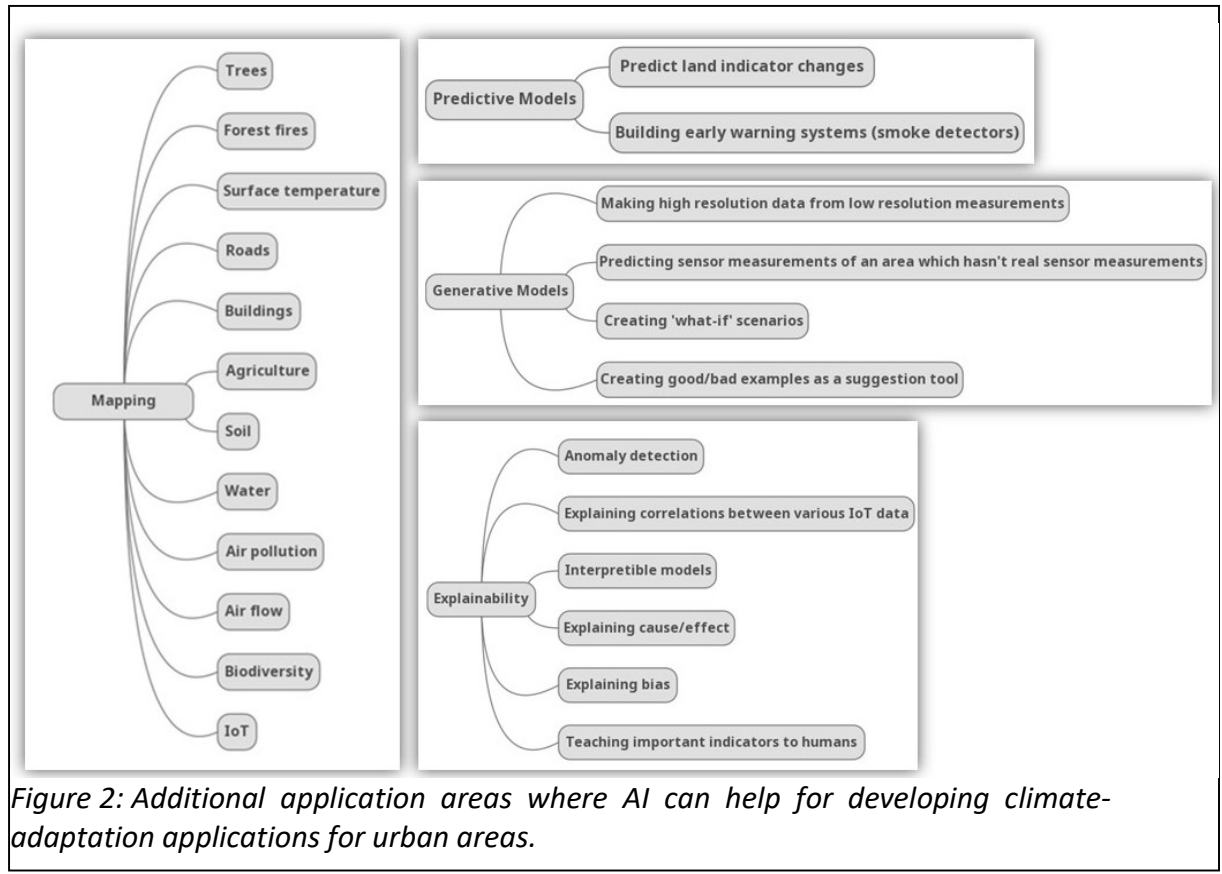

Figure 2: Additional application areas where AI can help for developing climate-adaptation applications for urban areas.

### 3.1.4. Explainability

It is possible to use explainable AI (XAI) and interpretable AI (also known as

'interpretability') methods to make the AI models and their results truly interpretable for general audience and provide further insights into the action goals of the policy makers [163]. As an example of using the interpretability methods in the context of the SDGs, in an early study Vinuesa and Sirmacek [162] illustrated that such interpretability methods could be used for tracking poverty in urban areas using satellite images and CNNs as developed by Jean et al. [82]. This work essentially identifies features such as night-light intensity, roofing material, distance to urban areas, etc., to predict the average economical consumption per capita and day. Vinuesa and Sirmacek [162] showed that adding interpretability to this model would help to really understand the influence of each parameter on the outcome, yielding a more robust and useful tool to track poverty and coordinate actions. In fact, the symbolic representation may help to understand which of these factors should be supported or suppressed to shift the poverty situation of the region to a better level.

**3.2. Non-intrusive evaluation of air quality in urban areas through AI**

When it comes to SDG 11 (on sustainable cities), we will focus on some relevant applications, including those aimed at extracting urban-development and environment-biodiversity indicators using fully-automated AI methods with remote sensing and other IoT data collected from smart cities. These indicators provide opportunities to: i) effectively monitor SDG 11 indicator11.3.1 on land-use efficiency; ii) observe the alignment of smart cities with other SDGs; iii) have early abnormality-detection possibilities when the indicators appear to be outliers; iv) better understanding of which urban development and which environmental indicators provide the best models for observation of the smart cities; v) create realistic scenarios to know which urban-development indicators make positive impact on the alignment of the smart city with the SDGs; and vi) create disaster scenarios to actually know and be prepared for the cases of observing unexpected indicator values. Another area where AI has great potential for SDG 11 is the development of robust non-intrusive-sensing methods to be able to more accurately determine the pollution levels and regions of extreme temperature in urban areas. It is important to note that around 90% of the population in the European Union (EU) were subjected to pollution levels exceeding those recommended by the World Health Organization (WHO) between 2014 and 2016 based on data by the European Environment Agency (EEA). It is estimated that these pollution levels produce around 800,000 premature deaths per year in the EU [101]. When it comes to extreme temperatures, the UHI phenomenon [110] mentioned above was connected with around 70,000 deaths in Europe during the summer of 2003 [67]. The great potential of AI in this context is further sup-ported by the fact that currently-available approaches for this are not accurate enough [24], and the EU is introducing the use of predictive models for pollutant-concentration measurements [1].Through flow prediction it is possible to provide, based on limited information, information about the temporal and spatial dynamics of the complete flow field (or certain relevant sub-sets of the field), up to a certain level of accuracy. One approach to perform the prediction is to first decompose the flow into spatial basis functions, such that only their temporal dynamics needs to be predicted. This can be accomplished by means of a well-known procedure, the proper-orthogonal decomposition (POD), which was introduced by Lumley in the context of turbulent flows [108]. This methodology basically decomposes the spatio-temporal velocity signal into spatial modes and temporal coefficients. Certain studies have considered variations of this technique, for instance the extended proper orthogonal decomposition (EPOD) [17], to perform predictions of the flow based on sparse pressure measurements [72]. In the EPOD framework, so-called extended velocity modes can be defined by combining information from the measured pressure and velocity signals, thereby allowing predicting the velocity field from the pressure readings. Certain properties of the EPOD were employed by Hosseini et al. [73] to predict the wake of a wall-mounted obstacle (representing a single simplified building) from pressure readings on its leeward side. Note that if all the possible extended modes are used for the reconstruction, the EPOD method is equivalent to a linear stochastic estimation (LSE) of the predicted quantity [17]. It is however important to note that the EPOD framework essentially considers a linear relationship between the measured and predicted quantities, which is insufficient to obtain accurate predictions given the complexity of the turbulent flow in urban environments. This was evaluated by Mokhasi et al.[116], who reached the conclusion that it is possible to obtain significantly better predictions of the temporal dynamics in such cases by using non-linear prediction methods.

So far we have discussed one approach to flow reconstruction mainly relying on first per-forming flow decomposition into spatial basis functions and then predicting the temporal evolution of the mode amplitudes via linear or non-linear methods. In some cases it is convenient to directly reconstruct the temporal evolution of the flow field (without a previous decomposition step) in certain regions of the domain: for instance, if we are interested in obtaining an accurate evolution of the flow on a certain horizontal or vertical plane. There have been some attempts to accomplish this type of reconstruction in the literature using linear methods. For instance, Illingworth et al. [78] employed a linear dynamical-system approach based on the resolvent-analysis framework [112] to predict the velocity field on a horizontal plane based on the velocity field from another horizontal plane, both of them being in the logarithmic region of a turbulent channel flow. On the other hand, other recent studies [43, 155] employed LSE to predict different horizontal planes of the flow in a turbulent channel based on wall measurements such as the pressure, and the two components of the wall-shear stress. Sasaki et al. [141] recently assessed flow-reconstruction methods based on single- and multiple-input linear transfer functions, which can then be used as convolution kernels to predict the fluctuations in a spatially-developing turbulent boundary layer. In particular, they performed predictions of the near-wall flow based on horizontal velocity fields in the outer region, and they also reconstructed the flow based on wall measurements. Note that the linear methods are able to provide only modest predictions close to the plane used as an input, and the accuracy of the reconstruction rapidly degrades farther away. This is because turbulent flows exhibit both linear (superposition) and non-linear (modulation) scale-interaction phenomena [38], therefore linear methods only provide an incomplete prediction. In fact, Sasaki et al. [141] also documented significant improvements in the predictions when using non-linear transfer functions to relate the input and the output.

Recent work by Guastoni et al.[59] reports a flow-reconstruction analysis in a turbulent open channel, where they predicted the turbulent fluctuations on different horizontal planes using the spatial distribution of the two wall-shear-stress components and the wall pressure. To this end, they employed a particular type of neural network, namely the so-called convolutional neural network (CNN) [98], which is widely used in computer vision. To summarize their results, close to the wall they were able to predict the streamwise fluctuation peak with less than 1% error, and farther away from the wall they obtained good results using a combination of a CNN and POD [59]. Despite the fact that this study was conducted in the context of turbulent channels, more complex geometries such as simplified urban environments [153, 165] can also be considered, including other quantities such as temperature and pollutant concentration. In fact, Güemes et al. [60] documented the potential of using GANs (which are discussed above) for predictions where few sparse measurements are available, and several additional studies have reported the possibility of using long-short-term memory (LSTM) networks for temporal predictions in turbulence [42, 151]. Consequently, deep neural networks are an excellent choice to predict horizontal (or vertical) sections of the flow field (as well as temperature and pollutant concentration) using wall data, thereby significantly improving currently available prediction techniques in urban flows. A schematic representation of the process is shown in Figure 3, and we argue that AI can certainly contribute towards the achievement of higher air quality in urban environments via sparse measurements. Also, to address the gap caused by the sparsely-distributed air-quality monitoring networks at the city level, Gupta et al. [63] and [62] proposed the simulated annealing based optimization method to capture data with higher precision at city level, with the opportunity to enable air-quality data collection more inclusive and encourage with citizen participation.

### 3.3. The role of AI in efficient and sustainable urban mobility

The urgent need of improving transportation within the urban environment is directly addressed in target 11.2 on affordable and sustainable transport systems. Moreover, an increased transportation efficiency, both from a sustainability and a connectivity perspective, would enable the achievement of targets 11.6 (reduced environmental impact of cities) and 11.a (implementation of urban and regional planning and increased inter-urban population integration). Transportation accessibility and efficiency has long been identified in the literature as a critical factor for social cohesion and inclusion [25, 131, 149], and the implementation of an integrated and inter-modal transportation system in urban and metropolitan areas is a key factor towards achieving a livable city [106, 166]. Furthermore, due to its current major dependency on internal combustion engines, transportation plays a big role in ambient air pollution, making it an area critical for any SDG-related intervention in the urban environment [46]. After a comprehensive literature review, we have identified three main areas in which AI can act as an enabling agent: guidance of urban transportation policy, urban-mobility planning and modelling and (connected) autonomous vehicles development.

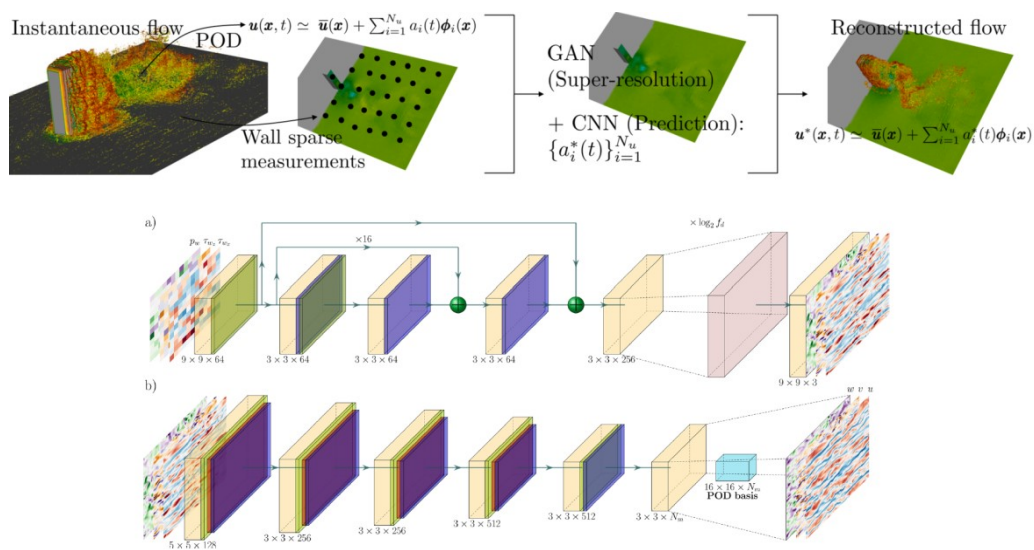

*Figure 3: Schematic representation illustrating (top) AI application to non-intrusive prediction of flow and pollutants in an urban environment; and (bottom) application of GANs*

*to super-resolution and prediction from sparse measurements in a turbulent flow. Bottom panel reprinted from Ref. [60], with permission of the publisher (AIP Publishing).*

## 4. AI for ambitious climate-action targets

In this section we will expand the evidence base and analysis on the role of AI in achieving SDG13 on climate action, as well as other broader objectives related to the climate crisis, including but not limited to the achievement of the Paris Agreement. A summary of the areas where AI can help to achieve SDG 13 is provided in Figure 4.

### 4.1. The potential role of artificial intelligence to combat climate change

Climate change could undermine the achievement of at least 72 targets across the SGDs, including outcomes for healthy and sustainable societies [119, 138]. Storms, droughts, fires, and flooding have become more frequent and stronger [49]. Global ecosystems are unstable, including the agriculture and natural resources on which humanity depends. The intergovernmental report on climate change in 2018 reported that the world would encounter catastrophic consequences unless global greenhouse gas emissions are removed within thirty years [7]. Yet year after year, these emissions rise. Addressing climate change includes mitigation (reducing emissions) and adaptation (preparing for unavoidable consequences). Both have multifaceted aspects. Mitigation of greenhouse-gas (GHG) emissions requires improvements in electricity systems, transportation, buildings, industry, and land use. Adaptation needs planning for resilience and disaster management, given an understanding of climate and extreme events. Artificial Intelligence (AI) has the potential to enhance global efforts to both mitigate GHG emissions and adapt re-quired planning [65, 163]. There is evidence that AI advances will support the understanding of climate change and the modelling of its possible impacts. AI is helpful in dealing with several climate-change mitigation measures. For example, AI can help to capture patterns and process temperature change data and carbon emissions[12,178], predict extreme weather events caused by climate change [48], recognize the effects of climate on health [15], understanding the energy needs and manage energy consumption [9, 91], monitoring impact in biodiversity due to climate change [41, 94], transforming transportation system for decreasing carbon emissions and making it more efficient in energy management and routing [8, 76, 114], monitoring impact on ocean [105], predict impacts for enabling precision agriculture [146], support in smart recycling [139], assist carbon capture and geo-engineering [113], and creating awareness about the climate impact [52]. AI will support low-carbon energy systems with high integration of renewable energy and energy efficiency, which are all needed to address climate change.

**4.2. AI in support of understanding climate change**

An extensive range of social areas are challenged by climate change. This fact demands remarkable adaptation to tackle future changes in weather patterns. AI has enhanced dramatically, provoking advancement in various research sectors, and also proposed as aiding climate analysis [137, 142]. AI can be integrated to discovered climate connections by the Earth System Model (ESM) to support improved warnings of approaching weather features, like extreme weather events. While ESM development is of principal importance, it is suggested that a parallel emphasis on implementing AI to understand far more on existing models and simulations [77]. AI advances will support the understanding of climate change and the modeling of its possible impacts. Therefore supporting adaptive capacity to climate change [158].AI techniques are utilized to investigate a tremendous amount of unstructured and heterogeneous data and disclose and extract complex and sophisticated relations among the data without the demand of an explicit analytical model of those relations [36, 70], supports in understanding the climate anomalies [181]. AI is advancing the way we understand the impact of climate change on biomass [177], hydrology [58], extreme events such as droughts [183] and support in taking mitigation measures [21, 54]. The application of AI techniques in extracting meaningful patterns and data sets from the rapidly increasing data deluge for the aim of coping with challenges related to the weather forecast, climate monitoring [54], and decade-wise prediction is inevitable [36].

Many AI techniques help identify inter-seasonal connections, linking potential climate-induced risk, aiding adaptation planning, e.g., timely crop sowing and mitigate natural disasters. Recent advancements such as drones and the IoT with AI support, improving the efficiency of existing systems by offering possibilities to extend mission coverage with refined spatial and temporal resolutions. A recent contribution to the domain includes the hybrid frameworks powered with deep-learning techniques to classify images aiding in natural disasters such as avalanches, cyclones, and fires [69, 120]. AI also plays a vital role in a wide range of responses to the climate crisis, mainly focused on mitigating existing emissions. Recent studies highlight the relevance of using AI in decreasing environmental emissions produced by industries and urban spaces [81, 88], and foster circular economy vision [11, 176].

## 4.3. AI in support of low-carbon energy systems

The obtainment of fuels and raw materials for the electricity grid, the process of generating and storing electricity, as well as the transmission of electricity to end-use consumers as called electricity system are responsible for around a quarter of human-caused greenhouse gas emissions each year [26]. Furthermore, since other energy-intensive sectors as buildings and transportation seek to replace GHG-emitting fuels, demand for low-carbon energy systems will grow. AI will contribute to rapid transition to low-carbon energy sources (like solar, wind, hydro, and nuclear) and decreasing the share of carbon-intensive sources (like natural gas, coal, and other fossil fuels). Renewable energy resources are appearing as sustainable alternatives to fossil fuels. They are much safer and cleaner than conventional fossil sources. With the remarkable advancements in technology, the renewable energy sector has made outstanding progress in the last decade [20]. However, there are still a wide variety of challenges associated with renewable energies that can be addressed with the help of innovative techniques. AI can analyze the past, optimize the present, predict the future, and digitalize the energy sector. The unpredictability of the available resource is one of the most significant challenges of producing renewable energy [66]. The electric grid is evolving rapidly with integrating variable renewable energy sources [143]. Due to the inherent intermittence of renewable energy sources, the current grid encounters many challenges in combining the diversity of renewable energy [66]. The utility industry requires intelligent systems to improve the integration of renewable energies with the existing grid and let renewable energies play an equal role in the energy supply. The energy grid collects a large amount of data by interconnecting with devices and sensors. A techniques could: i) systematically analyze a vast amount of data generated in plants; ii) translate the complex data into visualizations and insights that everyone can take advantage of; iii) discover, interpret, and communicate meaningful patterns in data; iv) diagnose and understand the reason behind the patterns in data in the past; v) predict what is most likely to happen in the future; vi) apply data patterns towards effective decision making; and finally vii) recommendations to be taken to affect the outcomes [103]. This AI-based data-driven information will give the grid planners and operators new insights to plan and operate the grid more efficiently [89]. It also offers flexibility to the energy providers to cleverly adjust the supply with demand [186].While the biggest goal of AI in renewable energy is to manage intermittency, it can also offer improved safety, efficiency, and reliability. It can help understand the energy consumption patterns and identify the devices' energy leakage and health [170].

However, AI could be used to identify technically-recoverable oil and gas resources and optimize the coal sector, reducing global fossil fuel prices and therefore reducing the competitiveness of renewable energy sources [102]. The application of AI in the oil and gas industry is very quickly enhancing. AI gradually gets through various stages of the oil and gas industry, such as intelligent drilling, intelligent extraction, intelligent pipeline, intelligent refinery, etc., and it will become the future research direction [102]. While AI can provide many advantages to the energy system, it can also cause some concerns like vulnerability to cyber-attacks, privacy and data ownership, and economic disruption [4]. Reported disruptions related to cyber-attacks in energy systems have been relatively small. Nevertheless, the increasing application of digitalized equipment and the growth of the internet of things (IoT) in energy systems could make cyber-attacks easier and cheaper to organize [4].

**4.4. AI in service of energy efficiency**

AI is believed to be the critical aspect in the energy systems, dealing with different energy practices (electricity, hydrogen-based fuels, wind, nuclear, solar, and other renewable sources, carbon capture) along with the end-use perspective (electrical appliances, transportation, heating, manufacturing, industry, and others) [57, 100, 136]. AI can help in underpinning energy diversity and localization for infrastructure planning, energy consumption forecasting, and intelligent controlling [32, 169]. AI techniques are used for improving the design, manufacturing, and optimization of energy-consumption-related aspects and also identifying the optimal materials, improving the safety of energy-use [39, 61, 150, 159]. AI could support the usage of Smart Grids, which in turn could increase the efficiency of local and global energy systems. There are two enabling factors for making a Smart Grid possible. In the first place, the deployment of modern IoT devices allows increasing the quantity and quality of data obtained from the network. Secondly, the big data collected can now be processed by AI to obtain quick results on decision-making that would be impossible for human operators [6]. The power grid is a complicated adaptive system under semi-autonomous distributed control with a lot of uncertainties. The integration of renewable energy such as solar and wind farms, electric and plug-in hybrid vehicles adds further complexity and challenges to different levels of the power grid. Many efforts have been put into smart grid development to coordinate theinterests of electric consumers, utilities, and environmentalists [160]. Real-time data in buildings and weather forecast, combined with smart systems, could predict when heating and cooling are needed, thus increasing system efficiency [4, 180]. AI could conduct the active demand-side management for households in smart grids, which contain distributed solar photo-voltaic generation and energy storage [37]. Smart demand response, for instance, could provide 185 Gigawatts (GW) of electricity system flexibility, approximately equal to the currently combined installed electricity supply capacity of Australia and Italy [4]. Consensus exist among experts globally that our the future energy supply should be economical, cleaner, and safer [56]. This in return will help in sustainable development by making electricity more affordable and accessible, decreasing GHG emissions, and efficient grid operations and reliable maintenance of power infrastructure, if used mindfully.

**4.5. Engagement of AI on climate change**

AI can reduce costs, increase productivity, raise resource intensity, and enhance efficient public services [163]. AI has been proposed as an enabler for new ambitious policy proposals for addressing climate change, such as being used for the implementation of personal carbon allowances [50]. However, there are also risks and downsides associated with AI that we all must be aware of being able to address any potential short-/long-term undesired impact [64, 65]. AI can have a significant impact on global energy demand. Developed AI technology, research, and product design may require extensive computational resources, which are only accessible through advanced computing centers. Recent studies on the energy demand and emissions associated with training and development of AI models have indicated the broader consequences of this rapid development [68]. Evidence is also emerging about the substantial climate impact of AI development [97]. These carbon footprints are associated mainly with the rapid development and training of AI algorithms with little consideration to the overall impact on the Earth system. Some estimates further show that the total electricity demand of ICT could grow up to 20% of the global electricity demand by 2030, from around 1% today [111].

With the increasing amount of data from diverse sources, the role of AI will steadily increase. In particular, AI will play a vital role considering the increasing debate on green, low-carbon electricity generation through optimal energy storage scenarios. Several efforts have been made to decrease the carbon footprint of data centers by investing in energy-efficient infrastructure and switching to renewable sources of energy [86, 111]. Considering the current state of the wide range of dependencies in one form or other on AI and associated services of AI systems (e.g. data collection and storage, hardware requirements and global shipments, training AI/ML models, etc.), uncertainty persists in realizing the comprehensive carbon footprint of AI. It is crucial to keep pace with the growing demand for AI infrastructure and whether efficiency gains by AI can be equally realized globally is an essential factor considering the environmental impact. The evidence to realize the net energy effects of AI and associated digital technologies are emerging. The indirect effects of using AI are likely to have a more considerable impact than the energy savings. The impact could be positive or negative depending on how mindfully it was utilized. Rebound and systemic effects are essential to be integrated to realize the complete picture of whether -or under which conditions and context- the AI services lead to a net positive or negative impact. Furthermore, the increased digitalization of strategic infrastructure exhibits some clear cyber-security challenges, increasing resource requirements along with time.

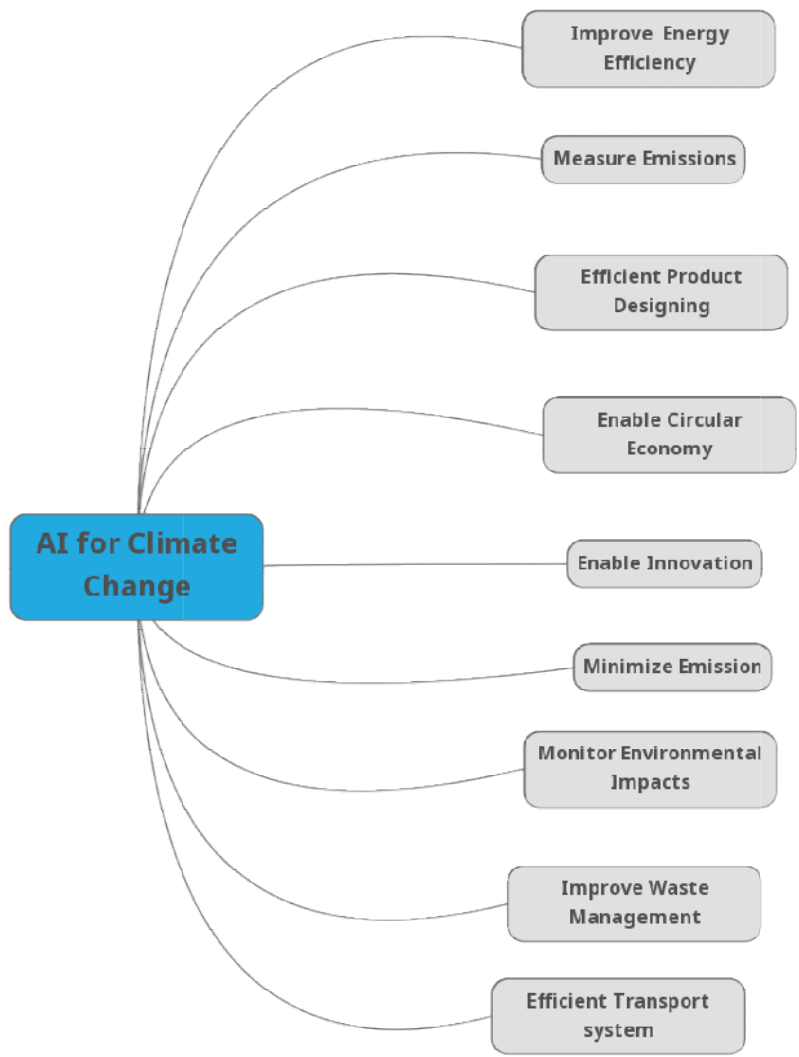

*Figure 4: Applications of AI that can help to achieve SDG 13.*

## 5. Conclusions and outlook

AI deployment has mayor consequences on society, the economy and the environment, and consequently on the SDGs. As evidenced by the COVID-19 crisis, AI can be a tool to increase the resilience of urban populations during times of crisis, but it also has also negative impacts. An understanding of these effects is essential so that we can tackle other important crises, such as the climate emergency. In this contribution we summarized the potential of AI to help achieve the SDGs related to healthy and sustainable societies, i.e. SDG 3 (on good health), SDG 11 (on sustainable cities) and SDG 13 (on climate action).When it comes to SDG 3, AI can help combat the shortage of health-care workforce, which affects greatly the low-and middle-income (LAMIC) countries. There is potential in the optimization of available resources through triaging and improved diagnosis, as well as in more detailed screening and prognosis. AI can also help in the context of automatic drug discovery, or GeoAI for patient-location history, which may help against epidemics and pandemics. This is aligned with the One Health approach. This of course has socio-ethical concerns, including privacy and data handling, increased inequalities due to lower AI literacy or access, and the need for specialized training of healthcare professionals. In healthcare-decision assistance, there are problems with biased training datasets, which may lead to under-representation of certain ethnicities or social groups. Overall, there is tremendous potential in the context of this SDG, as long as the possible pitfalls are understood and properly handled. Regarding SDG 11, AI can help with various aspects to understand the climate impact and to prepare adaptation strategies for urban areas. To this end, satellite imaging and IoT-sensor-based data-collection methods are often preferred, because of their capacity to provide sustain-able data in large areas over long periods of time. We discuss the use of AI for mapping and predictive/generative modeling, as well as the use of explainable-AI methods in order to provide solutions to understand vegetation cover, wildfire spreading, heat-island impacts, water security, air quality and other applications to support climate adaptation. Explainable-AI methods are not only found useful for understanding the climate indicators in more depth, but they are also found important for increasing trust on AI models by bringing more transparency on their functionality (thus avoiding black-box modeling).

Finally, prediction and pattern-recognition capabilities, which may help to better prepare for extreme weather events, monitoring biodiversity, and can provide improved climate modeling, are areas where AI can help to achieve SDG 13. Also, increased energy efficiency through integration of largely-varying renewable-energy sources into the energy mix, together with consumption forecasting and grid optimization (smart grids) are relevant areas fueled by AI. In this context, attention must be paid to cyber-security in AI-driven electrical grids, due to possible disruptions and data-privacy problems. Finally, it is important to note that there is a large car-bon footprint related with training complex and expensive AI models, and there is a strong need to decrease carbon footprint of the data centers used for model development. To conclude, the increased ability to acquire, process and analyze large amounts of heterogeneous data is the main driver behind the AI disruption. Pattern recognition, as well as the reconstruction and predictive capabilities of state-of-the-art AI models, present great opportunities in achieving healthy and sustainable societies. There are already a number of applications of AI related to health, smart cities and smart grids in use, proving its potential. Nevertheless, the increased complexity of these models, which (in the case of deep learning) essentially act as a black boxes consuming vast amounts of data, could hinder some of the efforts towards achieving the SDGs, in relation to equality and climate change. Privacy, data-management and governance, the carbon footprint associated with the training and deployment of AI models, as well as their interpretability are identified as key aspects which could be defining in the role of AI in achieving healthy and sustainable societies.

**Acknowledgments**

RV acknowledges the support of the KTH Sustainability Office and the KTH Digitalization Platform. SG acknowledges the support provided by the German Federal Ministry for Education and Research (BMBF) in the project "digitainable" .